\documentclass[prd,aps,showpacs,showkeys]{revtex4-1}
\usepackage{bm,amsfonts,latexsym,amsmath,amssymb,amsbsy,graphicx}

\newcommand{\bb}{\begin{eqnarray}}
\newcommand{\ee}{\end{eqnarray}}
\newcommand{\ba}{\begin{align}}
\newcommand{\ea}{\end{align}}

\begin{document}

\title{\bf  Polarization operator in the 2+1 dimensional quantum electrodynamics
with a nonzero fermion density in a constant uniform magnetic field}
\author{V.R. Khalilov \footnote{Corresponding author}}\email{khalilov@phys.msu.ru}
\affiliation{Faculty of Physics, M.V. Lomonosov Moscow State University, 119991,
Moscow, Russia}
\author{I.V. Mamsurov}
\affiliation{Faculty of Physics, M.V. Lomonosov Moscow State University, 119991,
Moscow, Russia}

\begin{abstract}
The polarization operator (tensor) for planar charged  fermions
in constant uniform magnetic field is calculated
in the one-loop approximation of the 2+1 dimensional
quantum electrodynamics (QED$_{2+1}$) with a nonzero fermion density.
We construct the Green function of the Dirac equation
with a constant uniform external magnetic field in the QED$_{2+1}$ at
the finite chemical potential, find the imaginary part of this Green function
and then obtain the polarization tensor related to the combined contribution from real
particles occupying the finite number of energy levels and magnetic field.
We expect that some physical effects under consideration seem
to be likely to be revealed in a monolayer graphene sample in the presence
of external constant  uniform magnetic field
$B$ perpendicular to it.
\end{abstract}



\maketitle

%\pacs{03.65.-w, 03.65.Pm, 03.65.Ge, 04.20.Jb}

\section{Introduction}

Planar charged fermions governed by Dirac equation  with external
electromagnetic fields attract considerable interest
in connection with problems of the quantum Hall effect \cite{2},
high-temperature superconductivity \cite{3} as well as graphene (see, e.g., \cite{6}
and \cite{7,8,9}). In graphene, the electron dynamics at low energies is described by
the two-dimensional Dirac equation for massless fermions \cite{7,10,11,12} though the case of massive charged fermions is also of interest \cite{13}.

It is well known \cite{11ll} that every energy level of planar electron in
an uniform magnetic field is degenerate and the number of degenerate states
per unit area is $|eB|/(2\pi\hbar c)$. So,  the kinetic energy of the
electrons can completely quenched at strong magnetic fields. Moreover,
the kinetic energy per the Dirac electron  is of order
$\varepsilon\approx v_F\sqrt{2|eB|\hbar c}$ (where $v_F$ is the dimensionless
Fermi-Dirac velocity)  and is comparable with the Coulomb energy per electron $E_C=e^2/(\epsilon_0 l_B)$, where $l_B=\sqrt{\hbar c/|eB|}$ is the so-called magnetic length and $\epsilon_0$ is
the dielectric constant of the medium \cite{9}. In the two-dimensional electron gas,
this enhancing of Coulomb interactions between  electrons in the presence of
strong magnetic fields, probably, leads to the fractional quantum Hall effect \cite{Laugh,goer}.

Important physical quantities related to the vacuum polarization
are the vacuum charge and current densities induced by the background field.
Polarization effects
in the massive QED$_{2+1}$  with a constant uniform magnetic field and
with a nonzero fermion density were studied in \cite{angs,kh0}.
In particular, the contribution of the induced Chern-Simons term
to the polarization tensor and the effective Lagrangian with the electron density
corresponding to the occupation of $n$ Landau levels in an uniform magnetic field
were calculated in \cite{kh0}.

Since the effective fine structure constant in graphene is large, the QED$_{2+1}$ effects
can be significant  already in the one-loop approximation.
Important quantum relativistic effects  were discussed
in \cite{kgn1} (Klein paradox) and \cite{casim} (Casimir effect).
The polarization operator in
a strong magnetic field perpendicular to the graphene membrane
has been calculated  in \cite{three}.
The problem of light absorption in graphene was investigated in \cite{ifdv}
and  the Faraday effect in a monolayer graphene sample in a strong constant  uniform
magnetic field  perpendicular to it was considered
in \cite{ifdv1}.

The  induced vacuum current in the field of a solenoid perpendicular to
the graphene sample was investigated in \cite{15},
and vacuum polarization in the QED$_{2+1}$ with an Aharonov--Bohm (AB) potential
for massive and massless fermions was studied in \cite{kh1}.
The vacuum electric current  due to to vacuum polarization in the AB potential for massive case
was observed in \cite{34a} in ``a quantum-tunneling system using two-dimensional ionic
structures in a linear Paul trap''.
Very important phenomenon - charged impurity screening in graphene due to the vacuum polarization by a Coulomb field - was investigated  in \cite{12,vp11a,as11b,kn11c,13a}.
Effect of spin on the dynamics of the two-dimensional Dirac oscillator in
the magnetic cosmic string background was considered in \cite{dirosc}.

In this work, we have calculated the polarization tensor of planar charged  fermions
in the presence of an external constant uniform
magnetic field in the one-loop approximation of the QED$_{2+1}$ at the finite chemical potential.
We have shown that one-loop polarization tensor
induces physical effects, which seem to be likely to be revealed in a monolayer
graphene sample in a strong constant uniform magnetic field aligned
perpendicularly to the sample.

We shall adopt the units where $c=\hbar=1$.

\section{Vacuum polarization by a constant uniform magnetic field in QED$_{2+1}$}

The polarization operator (PO) in a constant uniform magnetic field in QED$_{2+1}$ is diagonal
with respect to the photon three-momentum and in the momentum representation is determined by
\bb
\Pi^{\mu \nu}(p)=-ie^2\int\frac{d^3k}{(2\pi)^3}{\rm tr}[\gamma^{\mu}S^c(k,B)\gamma^{\nu}S^c(k-p,B)],
\label{po0}
\ee
where $S^c(k, B)$ is the causal Green function of the Dirac equation with a constant uniform magnetic field $B$ in the momentum representation.
In the coordinate representation the Green function $S^c(x^{\mu}-x^{'\mu}, B)$ of
the Dirac equation for a fermion of the mass $m$ and charge $e$ in an external constant uniform magnetic field in 2+1 dimensions satisfies equation
\bb
 (\gamma^{\mu}P_{\mu} - m)S(x^{\mu}-x^{'\mu}, B) = \delta^3(x^{\mu}-x^{'\mu}), \label{Dirac}
\ee
where $x^{\mu}=x^0, x^1, x^2\equiv t, x, y$ is the three-vector,
$P_\mu = -i\partial_{\mu} - eA_{\mu}$ is the generalized fermion momentum operator.
The Dirac $\gamma^{\mu}$-matrix algebra in 2+1 dimensions is known to be represented
in terms of the two-dimensional Pauli matrices $\sigma_j$
\bb
 \gamma^0= \sigma_3,\quad \gamma^1=i\tau\sigma_1,\quad \gamma^2=i\sigma_2, \label{1spin}
\ee
where the parameter $\tau=\pm 1$ can label two types of fermions in accordance with the
signature of the two-dimensional Dirac matrices \cite{27}; it can be applied to characterize two states of the fermion spin (spin "up" and "down") \cite{4}.
We take the magnetic field vector potential in the Cartesian coordinates in the Landau gauge   $A_0=0,\quad A_1=0,\quad A_2=Bx$, then the magnetic field is defined as $B=\partial_1A_2-\partial_2A_1\equiv F_{21}$, where $F_{\mu\nu}$ is the electromagnetic field tensor.

The positive-frequency Dirac equation solutions (the particle states) in the considered field corresponding to the energy eigenvalues (the Landau levels)
\bb
E_n^+=E_n\equiv \sqrt{m^2+2n|eB|},\quad n=0,1, \ldots ,
\label{Ll}
\ee
is given by \cite{kh0}
\bb
 \Psi^+(t,{\bf r}) =\frac{1}{\sqrt{2E_n}}
\left( \begin{array}{c}
\sqrt{E_n+m}U_n(z)\\
-{\rm sign}(eB)\sqrt{E_n-m} U_{n-1}(z)
\end{array}\right)\exp(-iE^+t+ip_2y), \label{three}
\ee
where the normalized functions $U_n(z)$ are expressed through the Hermite polynomials $H_n(z)$ as
$$
U_n(z)= \frac{|eB|^{1/4}}{(2^nn!\pi^{1/2})^{1/2}}e^{-z^2/2}H_n(z), z=\sqrt{|eB|}(x-p_2/eB)
$$
and $p_2$ is the eigenvalue because  $-i\partial_y\Psi^+(t,{\bf r})=p_2\Psi^+(t,{\bf r})$.
All the energy levels except the lowest level ($n=0$) with $\tau=1$ for $eB>0$ and $\tau=-1$ for $eB<0$  are doubly degenerate on spin $\tau=\pm 1$. This means that the eigenvalues of the fermion energy except the lowest level are actually spin-independent in the configuration under investigation. For definiteness, we consider the case where $eB<0$.
The negative-frequency Dirac equation solutions (the antiparticle states) corresponding to
negative energies  $E_n^-=-E_n$ can be constructed from (\ref{three})
by means of the charge-conjugation operation.

The exact expression for the free electron propagator
in an external magnetic field in 3+1 dimensions
was found for the first time by  Schwinger \cite{sch}. We can use
the Green function in the considered magnetic field obtained for
the 2+1 dimensions  in \cite{kh0} in the momentum representation in the form
\bb
S^c(p)=-\frac{i}{|eB|}\int\limits_{0}^{\infty}\frac{dz}{\cos z}\exp\left[\frac{iz}{|eB|}
\left(p_0^2-m^2-{\bf p}^2\frac{\tan z}{z}+i\epsilon\right)\right]\times \nonumber\\
\left[(\gamma^0p^0+m)\exp(i\sigma_3z)-\frac{(\gamma^1p^1+\gamma^2p^2)}{\cos z}\right],
\label{maggreen}
\ee
where $z=|eB|s$ and $s$ is the ``proper time".
The main properties and tensor structure of the PO can be obtained from the requirements of
relativistic and gauge invariance and also from the symmetry of the external field. In
the considered external field the PO must be diagonal with respect
to the ``photon" three-momentum and  depend only on three independent scalars, which can be constructed from the three-momentum $p^{\mu}$ and the tensor of external magnetic field $F^{\mu\nu}$:
\bb
p^2=(p^0)^2-{\bf p}^2, \quad p_{\mu}F^{\mu\nu}F_{\nu\rho}p^{\rho}\equiv B^2{\bf p}^2, \quad
F^{\mu\nu}F_{\mu\nu}\equiv 2B^2.
\label{inv}
\ee
We introduce the orthonormalized system of the three-vectors $l^{\mu}_i$
\bb
l^{\mu}_1=\frac{1}{\sqrt{{\bf p}^2}}(0,-p_2, p_1), \quad l^{\mu}_2=\frac{1}{\sqrt{p^2{\bf p}^2}}({\bf p}^2,p_0p_1,p_0p_2), \quad l^{\mu}_3=\frac{1}{\sqrt{p^2}}(p_0,p_1,p_2),
\label{vec0}
\ee
which satisfy relations
\bb
g_{\mu\nu}l^{\mu}_il^{\nu}_k=g_{ik},\quad \sum\limits_{i,k}g^{ik}l^{\mu}_il^{\nu}_k=g^{\mu\nu},\quad
-(g^{\mu\nu}-p^{\mu}p^{\nu}/p^2)=\sum\limits_{j=1,2}l^{\mu}_jl^{\nu}_j,
\label{rel0}
\ee
where $g^{\mu\nu}$ is the Minkowski tensor $g^{11}=g^{22}=-g^{00}=-1$, and the nonzero diagonal components of $g^{ik}$ are $g^{11}=g^{22}=-g^{33}=-1$.

The vectors $l^{\mu}_i, i=1,2$ are not eigenvectors of the PO; the PO eigenvalue corresponding to the eigenvector $l^{\mu}_3$ is equal to zero  due to the gauge invariance
\bb
p_{\mu}\Pi^{\mu\nu}=\Pi^{\mu\nu}p_{\nu}=0.
\label{gauge}
\ee

As a result of calculations, we find the PO in the fully transversal form \cite{kh0}
\bb
\Pi^{\mu\nu}(p, b)=\frac{i^{1/2}e^2}{8\pi^{3/2}b^{1/2}}\int\limits_0^{\infty}\frac{x^{1/2}dx}{\sin x}
\int\limits_{-1}^1 du[\Pi_1l^{\mu}_1l^{\nu}_1+\Pi_2l^{\mu}_2l^{\nu}_2+C\times ie^{\mu\nu\rho}p_{\rho}] \times \nonumber\\
\times \exp\left[\frac{ix}{b}\left(\frac{p_0^2(1-u^2)}{4}-\frac{{\bf p}^2\sin [x(1+u)/2]\sin [x(1-u)/2]}{x\sin x}-m^2+i\epsilon\right)\right], \quad b=|eB|,
\label{PO1}
\ee
where
\bb
\Pi_1=p_0^2(\cos ux - u\cot x \sin ux)-{\bf p}^2\frac{u\sin x \sin 2ux -2\cos ux+\cos x(1+\cos 2ux)}{\sin^2 x},\nonumber\\
\Pi_2=p^2(\cos ux - u\cot x \sin ux), \quad C=2m\cos ux
\label{pi1pi2c}
\ee
and $e^{\mu\nu\rho}$ is a fully antisymmetric unit tensor.
We note that $\Pi^{\mu\nu}(p)$ is not symmetric tensor in 2+1 dimensions.
In (\ref{PO1}) last term is the so-called  induced  Chern-Simons term;
 using the relation  $e^{\mu\nu\rho}p_{\rho}=-\sqrt{p^2}(l^{\mu}_1l^{\nu}_2-l^{\mu}_2l^{\nu}_1)$
it can be written in another form.
It should be noted that the mass term in the considered QED$_{2+1}$ model is not
invariant with respect to the operations of spatial (and time) inversion, therefore
the induced Chern-Simons term must be generated dynamically by the external magnetic field;
it contributes to the vacuum polarization only in the one-loop QED$_{2+1}$ approximation \cite{berlee}.

In the limit $eB=0$,  we obtain from (\ref{PO1})
\bb
\Pi^{\mu\nu}(p)=\frac{i^{1/2}e^2}{8\pi^{3/2}}\int\limits_0^{\infty}\frac{ds}{\sqrt{s}}
\int\limits_{-1}^1 du[(1-u^2)(g^{\mu\nu}p^2-p^{\mu}p^{\nu})-2ime^{\mu\nu\rho}p_{\rho}]  \exp\left[is\left(\frac{p^2(1-u^2)}{4}-m^2+i\epsilon\right)\right].
\label{POfree}
\ee
The PO (\ref{POfree}) is a function of only one scalar $p^2$ and its analytic properties can be studied in the complex $p^2$ plane. It is important that $\Pi^{\mu\nu}(p)$ is a real function on the negative real half axis $p^2<0$ what allows us to perform integrations in (\ref{POfree}) for the domain  $p^2<0$ and to obtain \cite{kh0}
\bb
\Pi^{\mu\nu}(p)=\frac{e^2}{4\pi}\left[(g^{\mu\nu}p^2-p^{\mu}p^{\nu})\left(\frac{4m^2+p^2}{p^2\sqrt{-p^2}}
\arctan\sqrt{\frac{-p^2}{4m^2}}-\frac{2m}{p^2}\right)-\right. \nonumber\\  \left.-ime^{\mu\nu\rho}\frac{4p_{\rho}}{\sqrt{-p^2}}
\arctan\sqrt{\frac{-p^2}{4m^2}}\right].
\label{PO2}
\ee
It should be noted that the free polarization operator in 2+1 dimensions was obtained in another form and without the Chern-Simons term in \cite{pofree}.

The singularities of  $\Pi^{\mu\nu}(p)$ lie on the positive real half axis of $p^2$ and the point $p^2=4m^2$ is the branch point (the threshold for the creation of  fermion pairs),
so $\Pi^{\mu\nu}(p)$ is an analytic function in the complex $p^2$ plane with a cut $[4m^2,\infty)$;
the  domains $p^2<0$ and $p^2>4m^2$ are physical domains, and the domain $0\leq p^2\leq 4m^2$ is nonphysical. The function $\Pi^{\mu\nu}(p)$ in the whole domain of $p^2$ can be obtained by the analytic continuation of (\ref{PO2}). In the domain $p^2>4m^2$ the free polarization operator gains
the imaginary part, which is on the upper edge of the cut:
\bb
{\rm Im}\Pi^{\mu\nu}(p)=-\frac{e^2}{4\pi}(g^{\mu\nu}p^2-p^{\mu}p^{\nu})
\frac{4m^2+p^2}{p^2\sqrt{p^2}}.
\label{PO2im}
\ee
The imaginary part of the PO has a discontinuity $2{\rm Im}\Pi^{\mu\nu}(p)$ in going across the cut.

We can find the polarization operator for charged massless fermions putting in the above
formulas $m=0$. In particular, the free polarization operator for the case $m=0$ is a real function on the negative real half axis $p^2<0$ and has the extremely simple form:
\bb
\Pi^{\mu\nu}(p,m=0)=\frac{e^2}{8}\frac{(g^{\mu\nu}p^2-p^{\mu}p^{\nu})}{\sqrt{-p^2}}.
\label{PO0r}
\ee
The free polarization operator (\ref{PO0r}) is transverse.
Now the  point $p^2=0$ is the branch point (the threshold for the creation of massless fermion pairs),
so $\Pi^{\mu\nu}(p, m=0)$ is an analytic function in the complex $p^2$ plane with a cut $[0,\infty)$.
In the domain $p^2>0$ $\Pi^{\mu\nu}(p, m=0)$ is pure imaginary
and on the upper edge of the cut has the form:
\bb
\Pi^{\mu\nu}(p,m=0)=-i\frac{e^2}{8}\frac{(g^{\mu\nu}p^2-p^{\mu}p^{\nu})}{\sqrt{p^2}}.
\label{PO0im}
\ee
In this form the polarization operator has been calculated in \cite{pofree,ifdv}.
In the condensed matter problems, the $\Pi^{00}(p, m=0)$ component of polarization tensor
is actual that, for example, at $p^2<0$ has the form
\bb
\Pi^{00}(p,m=0)=-\frac{e^2}{8}\frac{{\bf p}^2}{\sqrt{{\bf p}^2-p_0^2}}.
\label{POm=0}
\ee
This formula is in agreement with that obtained for graphene in the so-called random phase approximation  in \cite{jgfg} (see, also \cite{8,cp1}).

It is convenient to represent the polarization operator for the case $m=0$
in a weak constant uniform magnetic field as follows
\bb
\Pi^{\mu\nu}(p,eB,m=0)=\Pi^{\mu\nu}(p,m=0)+
e^2(eB)^2\left[\pi_1(p)\left(g^{\mu\nu}-\frac{p^{\mu}p^{\nu}}{p^2}\right) +\pi_2(p)l^{\mu}_1l^{\nu}_1\right],
\label{poweak}
\ee
where the first term is the PO $\Pi^{\mu\nu}(p,m=0)$ and
$\pi_1(p), \pi_2(p)$ are functions of only $p$. In particular, in the domain $p^2<0$ they can be estimated up to constants as
\bb
\pi_1(p)\approx C/(-p^2)^{3/2}, \quad \pi_2(p)\approx C_1{\bf p}^2/(-p^2)^{5/2},
\label{estim}
\ee
where $C, C_1\sim 0.2$. Since fermions are massless dimensionless factors $eB/p^2$ are built
with $|p|$ in place of $m$.

To calculate the main contribution in (\ref{PO1}) in a strong magnetic field let us rotate
the contour  of integration in $x$ on $-\pi/2$ to obtain
\bb
\Pi^{\mu\nu}(p, b)=\frac{e^2}{8\pi^{3/2}\sqrt{b}}e^{-{\bf p^2}/2b}(g^{\mu\nu}p^2-p^{\mu}p^{\nu})\int\limits_0^{\infty}\frac{x^{1/2}dx}{\sinh x}
\int\limits_{-1}^1 du(u\coth x\sinh x -\cosh ux)\times\nonumber\\
\times \exp\left[-\frac{x}{b}\left(\frac{p_0^2(1-u^2)}{4}-m^2\right)\right].
\label{POstr}
\ee
Neglecting the term $\sim p_0^2/b$ in the exponent we integrate (\ref{POstr}) in $u$
and obtain
\bb
\Pi^{\mu\nu}(p, b)=-\frac{e^2}{4\pi^{3/2}\sqrt{b}}e^{-{\bf p^2}/2b}(g^{\mu\nu}p^2-p^{\mu}p^{\nu})\int\limits_0^{\infty}\frac{dx}{\sqrt{x}}
\left(\frac{\coth x}{x} -\frac{1}{\sinh^2 x}\right) e^{-xm^2/b}.
\label{POstr1}
\ee
We see that the integral is maximum at $m=0$, so we can estimate it putting $m=0$ in the exponent.
But the integrand does not contain $b$ at $m=0$. Therefore,  the leading contribution in $b$ in (\ref{POstr1}) is proportional to $b^{-1/2}$
\bb
\Pi^{\mu\nu}(p, b)=\frac{e^2C}{4\pi^{3/2}\sqrt{b}}(g^{\mu\nu}p^2-p^{\mu}p^{\nu}), \quad C\sim 2.
\label{POstr2}
\ee
This result is in agreement with that obtained in \cite{ash}. Thus,
the polarization tensor  in QED$_{2+1}$ is very small ($\sim |eB|^{-1/2}$)
in a large constant uniform magnetic field. This feature of the PO in QED$_{2+1}$ essentially differs
from that in the massive QED$_{3+1}$, where $\Pi^{33}$ component plays a major role
and increases in a large magnetic field as \cite{lokh} $\Pi^{33}\sim |eB|/m^2$. Physically, this is because only electrons from the lowest Landau level couple to the longitudinal components of the photon at $|eB|\gg m^2$ \cite{ash,lokh,gmsh}.
It is seen that the vectors $l^{\mu}_j, j=1,2$  are spacelike if $p^2>0$,  but if $p^2=0$, then the only vector $l^{\mu}_1$ is still spacelike, whereas $l^{\mu}_2\to p^{\mu}/\sqrt{p^2}$.
It means that the ``first''  photon mode ($\sim l^{\mu}_1$) becomes almost free in a very strong magnetic field,  and the ``second'' photon mode ($\sim l^{\mu}_2$) therefore does not exist.

\section{Green function and polarization operator at a nonzero fermion density}

Now we construct the Green function to the Dirac equation
for the case of a nonzero fermion density (the finite chemical potential).
The Green function with the finite chemical potential can be obtained
from the Green function in the momentum representation(\ref{maggreen})
by shifting the variable $p^0$ as $p^0\to p^0+\mu+i\delta{\rm sign}(p^0)$,
where $\mu$ is the chemical potential.
\bb
 S^c(p, m, \mu)=-\frac{i}{|eB|}\int\limits_{0}^{\infty}\frac{dz}{\cos z}
\exp\left[\frac{iz}{|eB|}\left([p^0+\mu+i\delta{\rm sign} p^0]^2-m^2-{\bf p}^2\frac{\tan z}{z}+i\epsilon\right)\right]\times \nonumber\\
\times\left[(\gamma^0(p^0+\mu)+m)\exp(i\sigma_3z)-\frac{(\gamma^1p^1+\gamma^2p^2)}{\cos z}\right],
\delta\to +0.
\label{mgreenp}
\ee
If $\mu>m$, there are real particles occupying the Landau levels, if $\mu<m$,
then there are no real particles. We also assume $\mu>0$ without loss of generality.
The integration path passes below the singularities in the integrand in (\ref{mgreenp}) and the imaginary term  $i\delta{\rm sign} p^0$ is essential near the poles \cite{lifpit}; the Green function (\ref{mgreenp}) has the poles at the points $p^0=\pm E_n-\mu$ as well as
an imaginary part ${\rm Im} S^c(p,\mu)$ related to the presence of real charged fermions.
Rotating the integration path over $z$ in (\ref{mgreenp})
into the lower half-plane we can extend (\ref{mgreenp}) onto the whole complex plane of $p^0$
with the cuts $[p^0_+,\infty), [-p^0_-,-\infty)$ at the real axis. Then, as a function of $p^0$ the Green function (\ref{mgreenp}) is a limit of some analytic function $S^c(p^0)$. Denoting the integrand in (\ref{mgreenp}) as $S(z)$, we represent
${\rm Im} S^c(p,\mu)$ via the discontinuities $\Delta S$ at the edges of the cuts in the form
$\int\limits_{-\infty}^{\infty} S(z)dz$. For this, we apply method for calculation of the PO discontinuities in the presence of various external electromagnetic fields
in vacuum \cite{sh,lokh} that was extended on the case of a nonzero fermion density  in \cite{kh0}.
 Finally, we obtain  $S^c(p,\mu)$ in the form
\bb
 S^c(p, m, \mu)=-\frac{i}{|eB|}\int\limits_{0}^{\infty}\frac{dz}{\cos z}
\exp\left[\frac{iz}{|eB|}\left([p^0+\mu+i\delta{\rm sign} p^0]^2-m^2-{\bf p}^2\frac{\tan z}{z}+i\epsilon\right)\right](\gamma P+M)-\nonumber\\
-2i\pi(\gamma \bar P+\bar M), \quad \delta\to +0.\phantom{mmmmmmmmmmmmmmmm}
\label{mgreenp1}
\ee
where
\bb
P^0=(p_0+\mu)\cos z +im\sin z, \quad M=m\cos z+i(p_0+\mu)\sin z,\quad {\bf P}=-{\bf p}/\cos z,\phantom{mmmmmm}\nonumber\\
\bar P=[m_0(p_0+\mu+m)/2m]\delta[(p_0+\mu-m)/m_0]e^{-{\bf p}^2/|eB|} + \phantom{mmmmmmmmmmmm}\nonumber\\  +mm_0\sum\limits_{n=1}^{\infty}\frac{(-1)^n}{E_n}I_{nn}({\bf p}^2/|eB|)\left[\delta[(p_0+\mu-E_n)/m_0]+\delta[(p_0+\mu+E_n)/m_0]\right],
 \phantom{mmmmmm}\nonumber\\
\bar M = [m_0(p_0+\mu+m)/2m]\delta[(p_0+\mu-m)/m_0]e^{-{\bf p}^2/|eB|}+\phantom{mmmmmmmmmm}\nonumber\\
+(p^0+\mu)m_0\sum\limits_{n=1}^{\infty}\frac{(-1)^n}{E_n}I_{nn}({\bf p}^2/|eB|)\left[\delta[(p_0+\mu-E_n)/m_0]+\delta[(p_0+\mu+E_n)/m_0]\right]\phantom{mmmmm}.
\label{imterm}
\ee
Here $m_0$ is the parameter of the dimension of mass, the Laguerre function $I_{nn}(x)=(1/n!)e^{-x/2} L_n(x)$, $L_n(x)$ is the Laguerre polynomial  and all the differences must satisfy inequalities: $m+\mu>0, E_n^+-\mu>0$, and $-m-\mu<0, E_n^--\mu<0$.

The Green function for charged massless fermions is easily derived from Eqs. (\ref{mgreenp1}), (\ref{imterm}) to read
\bb
 S^c(p, \mu, m=0)=-\frac{i}{|eB|}\int\limits_{0}^{\infty}\frac{dz}{\cos z}
\exp\left[\frac{iz}{|eB|}\left([p^0+\mu+i\delta{\rm sign} p^0]^2-{\bf p}^2\frac{\tan z}{z}+i\epsilon\right)\right](\gamma P+M)-\nonumber\\
-2i\pi(\gamma \bar P+\bar M), \quad \delta\to +0.\phantom{mmmmmmmmmmmmmmmm}
\label{mgreenpm=0}
\ee
where
\bb
P^0=(p_0+\mu)\cos z, \quad M=i(p_0+\mu)\sin z,\quad {\bf P}=-{\bf p}/\cos z,\phantom{mmmmmmmmmmmmmmmm} \nonumber\\
\bar P=[m_0(p_0+\mu)/2m]\delta[(p_0+\mu)/m_0]|_{m\to 0}e^{-{\bf p}^2/|eB|}, \quad
\bar M=[m_0(p_0+\mu)/2m]\delta[(p_0+\mu)/m_0]|_{m\to 0}e^{-{\bf p}^2/|eB|}+\phantom{mmmmmmm}\nonumber\\ +m_0(p^0+\mu)\sum\limits_{n=1}^{\infty}\frac{(-1)^n}{\varepsilon_n}I_{nn}({\bf p}^2/|eB|)\left[\delta[(p_0+\mu-\varepsilon_n)/m_0]+ \delta[(p_0+\mu+\varepsilon_n)/m_0]\right], \quad \varepsilon_n^{\pm}=\pm \sqrt{2|eB|n}.\phantom{mmmmmmmmmm}
\label{imtermm=0}
\ee
A massless fermion does not have a spin degree of freedom in 2+1 dimensions \cite{jacna} but
 the Dirac equation  for charged massless fermions in an external magnetic field
in 2+1 dimensions keeps the spin parameter. Therefore, all the energy levels except the lowest level $n=0$ are doubly degenerate; the levels with $n=n_r, \tau=1$  and $n=n_r+1, \tau=-1$ (where $n_r=0,1,2 \ldots$ is the radial quantum number) coincide.

In graphene, the set of Landau levels in an uniform magnetic field aligned
perpendicularly to the monolayer sample is given by (see, for example, \cite{9,goer})
\bb
\varepsilon_n=\pm v_F\sqrt{2|eB|n}, \quad n=0, 1, 2, \ldots,
\label{engrap}
\ee
where $v_F$ is the Fermi-Dirac velocity, and the $\pm$ signs
label the states of positive (electron) and negative (hole) energy, respectively.
They play the same role as the band index for the conduction (+)
and the valence (-) band. In addition, the states of positive
and negative energy for charged massless fermions has the same energy
$\varepsilon_0=0$ but opposite spins in the ground states.

We now discuss briefly the polarization tensor (PT) related to contributions coming
from real particles. We note that the  PT  contains terms with
$\mu \neq 0,  B=0$ and $\mu \neq 0,  B \neq 0$. Terms with
$\mu \neq 0,  B=0$ are very cumbersome and we do not give them.
In the absence of magnetic field  the $\Pi^{00}(p, \mu)$ component
of the PT in monolayer graphene  has been studied in one-loop approximation
in \cite{egvgvm,pkp,aqra}.
The $\Pi^{00}(p, \mu)$ component mainly contribute to the PT
in the presence of a weak magnetic field and so it should be taken into account in this case.

Here we give only the expression for the PT related to the combined contribution
from real particles and magnetic field. It is  natural to consider that
the medium contains particles, which implies $\mu>m$ and real antiparticles
are absent. We also assume that $\mu\neq m, E_n$, which means that all
the $n$ Landau levels are fully filled and no levels are partly filled.

As a result of long calculations (see, \cite{kh0}), one can obtain the PT related to the above combined contribution in the form
\bb
\Pi^{\mu\nu}(p,\mu b)=\frac{e^2}{2\pi|m|}\int\limits_0^{\infty}dx[p^2\Pi_rl^{\mu}_2l^{\nu}_2+mC_r\times ie^{\mu\nu\rho}p_{\rho}] \times \nonumber\\
\times \exp\left[ix\left(\frac{(p_0+\mu)^2}{b}-1\right)- \frac{{\bf p}^2(1+i\sin 2x-\cos2x)}{2b}-\epsilon x\right],
\label{POrp}
\ee
where
\bb
\Pi_r=\frac12 \sin\frac{2m(p_0+\mu)x}{b} +\frac{m}{E_n}L_n\left(\frac{{\bf p}^2\sin^2x}{2b}\right)
\sin\frac{2E_n(p_0+\mu)x}{b}\left[\frac{\mu^2-m^2}{2|eB|}\right] \nonumber\\
 C_r=\frac{i}{2} \cos\frac{2m(p_0+\mu)x}{b} +iL_n\left(\frac{{\bf p}^2\sin^2x}{2b}\right)
\cos\frac{2E_n(p_0+\mu)x}{b}\left[\frac{\mu^2-m^2}{2|eB|}\right]
\label{pr1pr2c}
\ee
Here the first terms give the contribution from $n=0$ Landau level  and $[(\mu^2-m^2)/2|eB|]\equiv N$ denotes the integer part of the function $u=(\mu^2-m^2)/2|eB|$, i.e. the largest integer $\le u$.

The part of the polarization tensor determined by the function $\Pi_r$ is of main interest. Calculating it for the case of  weak constant uniform magnetic field $(p_0+\mu)^2\gg |eB|$, we obtain
\bb
\Pi^{\mu\nu}(p,\mu b)=\frac{e^2}{2\pi}p^2\Pi_rl^{\mu}_2l^{\nu}_2.
\label{POrpw}
\ee
Here
\bb
\Pi_r=\frac{|eB|}{(p_0+\mu)[(p_0+\mu)^2-4m^2]}\theta(\mu-|m|)+
\frac{2|eB|}{(p_0+\mu)[(p_0+\mu)^2-4E_N^2]}\times N\theta(\mu-E_N^+),
\label{prw0N}
\ee
where $\theta(z)$ is the Heavyside function; therefore the first term gives the contribution
if the ground Landau level is occupied and the second one contributes  if the $N$ Landau levels are occupied.
If the magnetic field is strong $(p_0+\mu)^2\ll |eB|$, then
only the ground Landau level is occupied in the massive case and  we obtain
\bb
\Pi_r=\frac{p_0+\mu}{|eB|}, \quad \mu>m, \quad |eB|\gg m^2.
\label{prs0}
\ee
For the case of massless charged fermions we must put $m=0$ and  $E_N^+$ replace by $\varepsilon_N^+$ in Eq. (\ref{prw0N}). In a strong magnetic field $(p_0+\mu)^2/|eB|\ll 1$, we also obtain
\bb
\Pi_r=\frac{p_0+\mu}{|eB|}\theta(\mu) +\frac{2(p_0+\mu)}{|eB|}\times N\theta(\mu-\varepsilon_N^+),
\label{prsN}
\ee
where the first and second terms, respectively, give the contributions
if the ground Landau level is occupied and  the $N$ Landau levels are occupied.
The main feature is that the polarization tensor (\ref{POrp}) with a nonzero fermion density is extremely small ($\sim |eB|^{-1}$) in a strong constant uniform magnetic field.

\section{Discussion}

We have calculated the polarization tensor
in the one-loop approximation of the 2+1 dimensional
quantum electrodynamics  with a nonzero fermion density
in a constant uniform magnetic field. The polarization tensor  contains
the contributions from virtual (vacuum) and real charged particles
occupying the finite number of the Landau levels.
In particular, we have found that the polarization tensor in QED$_{2+1}$
in a strong constant uniform magnetic field is proportional to $\sim e^2 |p^2|/\sqrt{|eB|}, |eB|\gg |p^2|$ (contribution of virtual particles)
and $\sim e^2 |{\bf p}^2|(p_0+\mu)/|eB|, |eB|\gg (p_0+\mu)^2$  (contribution of real particles).
It means that  photons  become almost free in a very strong magnetic field.

It should be noted that if one, for instance, needs to investigate the propagation of
electromagnetic waves in a real graphene strip we must
consider it as  the three-dimensional object of extremely small
but nonzero thickness (see, also, \cite{ocbm}).

In addition, we emphasize that the polarization tensors (\ref{PO1}) and (\ref{POrp}) at nonzero
fermion mass are finite in the limit $p_{\rho}\to 0$ because they contain antisymmetric terms
that do not vanish in this limit. The coefficient multiplying $ie^{\mu\nu\rho}p_{\rho}$
in (\ref{PO1}) and (\ref{POrp}) with $p_{\nu}=0$ is called the induced  Chern-Simons coefficient and
its appearance in the effective Lagrangian of QED$_{2+1}$ with an external magnetic
field means that photons dynamically gain ``masses''. The induced  Chern-Simons
coefficient is calculated exactly and has the form
$$
C_{CS}=-{\rm sign (m)}\frac{e^2}{4\pi}\theta(|m|-\mu)+{\rm sign (eB)}\frac{e^2}{2\pi}\left(\frac12 \theta(\mu-|m|)+N\theta(\mu-E_N^+))\right),
$$
where the first term gives the contribution of virtual fermions.

%The work was supported in part by the Ministry of Education and Science of the Russian Federation
%grant (Agreement No 14.576.21.0025 of 27.07.2014).

\end{document}